\documentclass [12pt]{article}

\usepackage     {amsthm}
\usepackage     {graphicx}
\usepackage     {mathrsfs}
\usepackage     {amsfonts}
\usepackage     {makeidx}         
\usepackage     {multicol}        

\usepackage     {graphicx}
\usepackage     {mathrsfs}
\usepackage     {amsfonts}
\usepackage     {makeidx}         
\usepackage     {multicol}        

\begin{document}

\begin{center}
{\bf \Large Interaction of ``rigid'' quantum systems}

\medskip \medskip

{\large A.A. Kolpakov (Universit\'e de Fribourg, Suisse) \\
\medskip

A.G. Kolpakov (Marie Curie Fellow, Novosibirsk, Russia) }
\end{center}

\medskip
\medskip

We introduce the notion of a ``rigid'' quantum system as a system with constant relative positions of its nuclei and constant relative distribution of the electrons with respect to the nuclei. 

In accordance with this definition, a molecule which does not interact with other objects, is a ``rigid'' quantum system. Molecule is also ``rigid'' if it interacts with other objects, but the interaction does not change the intrinsic structure of the molecule (or this change can be neglected).

Several ``rigid'' quantum systems interact one with another in the quantum manner. The interaction is ruled by the Schr{\"o}dinger equation [1] written for all the particles of the systems under consideration.  We consider the case when the external potential is zero. 

Since the Schr{\"o}dinger equation for particle systems is invariant with respect to translations and rotations [1], every ``rigid'' quantum system can move as a ``rigid body'' without any change of its relative characteristics, in particular, without any change of its total energy. However, the total energy of several interacting ``rigid'' quantum systems changes when one of them moves even as a rigid body, because in quantum mechanics systems interact  always.

In reality, the above assumptions mean that the relative positions of nuclei and the distribution of electrons in every ``rigid'' quantum system changes so slightly, that the changes may be neglected. 

Our aim is to demonstrate that the interaction of two ``rigid'' quantum system (i.e., systems described by the Schr{\"o}dinger equation) can be expressed as an electrostatic interaction of two systems of charges. 

Although we demonstrate that the systems under consideration, finally, interact in the electrostatic manner, we cannot use the electrostatic model as a starting point, because a system of charges is electrostatic unstable [2]. We have to use the quantum model as a starting point, because only this model is in agreement with the fact of stable charged particles systems existence. Accepting the quantum model as a starting point, we have to derive the electrostatic principle for the interaction of charged systems from it.

In other words, we demonstrate that the following decomposition is possible:

1. the individual (microscopic) integrity and stability of every quantum system is a result of quantum forces (electrostatic forces cannot form stable objects);

2. the interaction of ``rigid'' quantum systems as individual (macroscopic) objects is powered by electrostatic forces corresponding to the (microscopic) distribution of charges in every  quantum system.

This decomposition is inspirited by the behaviour of some kinds of mo\-le\-cule, which may interact as individual objects without forming a new substance. 

Let one system $D_1$ be formed by $n_1$ electrons numbered $1,...,n_1$ and $m_1$ nuclei numbered $1,...,m_1$. Let another  system $D_2$ be formed by $n_2$ electrons numbered $n_1+1,...,n_1+n_2$ and $m_2$ nuclei numbered $m_1+1,...,m_1+m_2$.

We assume that the electric charge of each electron is $1$, and denote the electric charge of the $n$-th nucleus by $Z_n$ (in the electron units).

We use the notation
$$\stackrel{-}{{\bf x}}= ({\bf x}_1,...,{\bf x}_{n_1}) ,$$
$$\stackrel{=}{{\bf x}}=  ({\bf x}_{n_1+1},..., {\bf x}_{n_1+n_2}) .$$
With this notations ${\bf x} = (\stackrel{-}{{\bf x}}, \stackrel{=}{{\bf x}}) $.

The indices of the electrons are split into two sets: $I_1$ -- the indices of the electrons that belong to the first system and $I_2$ -- the indices of the electrons that belong to the second system. The indices of the nuclei are also split into two sets: $N_1$ -- the indices of the electrons that belong to the first system and $N_2$ -- the indices of the electrons that belong to the second system.

The problem could be written as [1]
\begin{eqnarray}   &  \displaystyle
 E(\phi) \to \min, \label{(1aaa)}
 \\  &  \displaystyle
 \int_{R^{3n_1}\times R^{3n_2}}  |\phi({\bf x})|^2 d{\bf x} = 1.
 \label{(1bbb)}
\end{eqnarray}
Arguments of the wave function $\phi({\bf x})$ is ${\bf x} = ({\bf x}_1,...,{\bf x}_{n_1}, {\bf x}_{n_1+1},..., {\bf x}_{n_1+n_2}) $.

The total energy $E(\phi))$ of the integral system (the system formed by both, the first and the second systems) is written as in [1] (we assume that there is no external potential):
\begin{eqnarray}   &  \displaystyle
 E(\phi)=\int_{R^{3n_1}\times R^{3n_2}}  \sum_{k \in I_1 \cup I_2} \Bigg[ |\nabla_k  \phi({\bf x})|^2 +
  \label{(1ccc)} \\[8pt] &  \displaystyle
 + \sum_{j\in I_1 \cup I_2} \frac{\phi^2({\bf x})}{|{\bf x}_k - {\bf x}_j|}
 + \sum_{n \in N_1 \cup N_2} \frac{\phi^2({\bf x})Z_n}{|{\bf x}_k - {\bf x}_n|} \Bigg] d{\bf x} +
 \nonumber \\[8pt] &  \displaystyle
 + \sum_{n <   m \in N_1 \cup N_2} \frac{Z_n Z_m}{|{\bf x}_n - {\bf x}_m|}.
 \nonumber
 \end{eqnarray}

The first three terms in (\ref{(1ccc)}) corresponds to the electrons: the first is the kinetic energy of the electrons, the second one is the potential energy of the mutual electrons interaction, the third one is the potential energy of the interaction between the electrons and the kernels. The last term in (\ref{(1ccc)}) is the potential energy of the mutual kernel interaction, the kinetic energy of the kernels is neglected (the notation $n < m \in M$ means that $n,m \in M$ and $n < m$).

In each of the ``rigid'' quantum systems, the electron density distribution is its own specific property, that does not depend on its counterpart. Thus, the electron density distributions over two ``rigid'' quantum systems are independent functions. So we can construct a solution to the problem (\ref{(1aaa)}), (\ref{(1bbb)}) for two ``rigid'' quantum systems in the form (which is often used in the analysis of the interaction of a pair of quantum objects [1]): 
\begin{eqnarray}   &  \displaystyle
\phi({\bf x}) = \stackrel{-}{\Phi} (\stackrel{-}{{\bf x}}) \stackrel{=}{\Phi} (\stackrel{=}{{\bf x}}).
 \label{(2)}
\end{eqnarray}

The advantage of the presentation used in (\ref{(2)}) is that if
$$ \int_{R^{3n_1}} |\stackrel{-}{\Phi} (\stackrel{-}{{\bf x}})|^2d\stackrel{-}{{\bf x}} =1, \,\,\, 
\int_{R^{3n_2}} |\stackrel{=}{\Phi} (\stackrel{=}{{\bf x}})|^2 d\stackrel{=}{{\bf x}} =1$$
than the equality
$$ \int_{R^{3n_1}\times R^{3n_2}} |\phi ({\bf x})|^2d{\bf x} =1 $$ 
is satisfied automatically.

By substituting (\ref{(2)}) into (\ref{(1ccc)}), we obtain
\begin{eqnarray}   &  \displaystyle
E=\sum_{k \in I_1 } \int_{R^{3n_1}}  |\nabla_k  \stackrel{-}{\Phi} (\stackrel{-}{{\bf x}})|^2 d\stackrel{-}{{\bf x}} \int_{R^{3n_2}} |\stackrel{=}{\Phi} (\stackrel{=}{{\bf x}})|^2 d\stackrel{=}{{\bf x}} +
\label{5} \\ &  \displaystyle
+ \sum_{k \in I_2 } \int_{R^{3n_1}} |\stackrel{-}{\Phi} (\stackrel{-}{{\bf x}})|^2  d\stackrel{-}{{\bf x}} \int_{R^{3n_2}} |\nabla_k  \stackrel{=}{\Phi} (\stackrel{=}{{\bf x}})|^2  d\stackrel{=}{{\bf x}} +
\nonumber \\ &  \displaystyle
 + \sum_{j\in I_1 k \in I_1} \int_{R^{3n_1}} \frac{\stackrel{-}{\Phi}^2 (\stackrel{-}{{\bf x}})}{|{\bf x}_k - {\bf x}_j|}d\stackrel{-}{{\bf x}}  \int_{R^{3n_2}} \stackrel{=}{\Phi}^2 (\stackrel{=}{{\bf x}}) d\stackrel{=}{{\bf x}} +
\nonumber \\ &  \displaystyle
+ \sum_{j\in I_1 k \in I_2} \int_{R^{3n_1}} \int_{R^{3n_2}} \frac{\stackrel{-}{\Phi}^2 (\stackrel{-}{{\bf x}}) \stackrel{=}{\Phi}^2 (\stackrel{=}{{\bf x}})}{|{\bf x}_k - {\bf x}_j|} d\stackrel{-}{{\bf x}} d\stackrel{=}{{\bf x}} +
\nonumber \\ &  \displaystyle
 + \sum_{j\in I_1 n \in N_1} \int_{R^{3n_1}} \frac{\stackrel{-}{\Phi}^2 (\stackrel{-}{{\bf x}})Z_n}{|{\bf x}_k - {\bf x}_n|} d\stackrel{-}{{\bf x}}  \int_{R^{3n_2}} \stackrel{=}{\Phi}^2 (\stackrel{=}{{\bf x}}) d\stackrel{=}{{\bf x}} +
\nonumber \\ &  \displaystyle
+ \sum_{k\in I_1 n \in N_2} \int_{R^{3n_1}} \int_{R^{3n_2}} \frac{\stackrel{-}{\Phi}^2 (\stackrel{-}{{\bf x}}) \stackrel{=}{\Phi}^2 (\stackrel{=}{{\bf x}})Z_n}{|{\bf x}_k - {\bf x}_n|} d\stackrel{-}{{\bf x}} d\stackrel{=}{{\bf x}} +
 \nonumber \\[8pt] &  \displaystyle
 + \sum_{n <  m \in N_1 \cup N_2} \frac{Z_n Z_m}{|{\bf x}_n - {\bf x}_m|}.
 \nonumber
\end{eqnarray}

We investigate the terms forming the right hand part of (\ref{5}). 

We have that 
\begin{eqnarray}   &  \displaystyle \label{is1}
\int_{R^{3n_1}} \stackrel{-}{\Phi}^2 (\stackrel{-}{{\bf x}})  d\stackrel{-}{{\bf x}}= 
\int_{R^{3n_2}} \stackrel{=}{\Phi}^2 (\stackrel{=}{{\bf x}}) d\stackrel{=}{{\bf x}} =1.
 \end{eqnarray}
 
We integrate in
 \begin{eqnarray}   &  \displaystyle
\int_{R^{3n_1}} \frac{\stackrel{-}{\Phi}^2 (\stackrel{-}{{\bf x}})}{|{\bf x}_k - {\bf x}_j|}d\stackrel{-}{{\bf x}}  
\end{eqnarray}
with respect to all variables $\stackrel{-}{{\bf x}}$ except for ${\bf x}_k$. As a result, we obtain
\begin{eqnarray}   &  \displaystyle 
\int_{R^{3n_1}} \frac{\stackrel{-}{\Phi}^2 ( \stackrel{-}{{\bf x}})}{|{\bf x}_k - {\bf x}_j|}
d\stackrel{-}{{\bf x}}   =
\int_{R^3} \frac{ \displaystyle \int_{R^{3n_1-3}}\stackrel{-}{\Phi}^2 (\stackrel{-}{{\bf x}}) 
d(\stackrel{-}{{\bf x}} \setminus {\bf x}_k)}
{\displaystyle |{\bf x}_k - {\bf x}_j|}d{\bf x}_k.
\end{eqnarray}

The symbol $\displaystyle \int_{R^{3n_1-3}}f (\stackrel{-}{{\bf x}}) d(\stackrel{-}{{\bf x}} \setminus {\bf x}_k)$ means the integration of the function $f (\stackrel{-}{{\bf x}})$ of the variables $\stackrel{-}{{\bf x}}= ({\bf x}_1,...,{\bf x}_{n_1}) $ with respect to all variables $\stackrel{-}{{\bf x}}$ except for ${\bf x}_k$ ($k \in 1,...,n_1$).

The symbol $\displaystyle \int_{R^{3n_1-3}}f (\stackrel{=}{{\bf x}}) d(\stackrel{=}{{\bf x}} \setminus {\bf x}_k)$ means the integration of the function $f (\stackrel{=}{{\bf x}})$ of the variables $\stackrel{=}{{\bf x}}$ with respect to all variables $\stackrel{=}{{\bf x}}$ except for ${\bf x}_k$ ($k \in n_1+1,...,n_1+n_2$).

\medskip 
The integral
\begin{eqnarray}   &  \displaystyle 
\int_{R^{3n_1-3}}|\stackrel{-}{\Phi} (\stackrel{-}{{\bf x}})|^2 d(\stackrel{-}{{\bf x}} \setminus {\bf x}_k)
\end{eqnarray}
has the meaning of the probability density of the event ``the $k$-th electron is placed in the vicinity of the point ${\bf x}$'' and can be regarded as the contribution of the density of the $k$-th electron to the density of the total electron field in the system $D_1$. We denote it by $\rho_{1k}({\bf x})$. By analogy, we denote the density of the electron field in the system $D_2$ by $\rho_{2k}({\bf x})$.

The sum over the indices of all electrons in the system $D_1$
\begin{eqnarray}   \label{rho} &  \displaystyle 
\rho_{1}({\bf x})=\sum_{j\in I_1} \rho_{1j}({\bf x})
\end{eqnarray}
is the density of the total electron field in the system $D_1$ at the point ${\bf x}$. A similar equality is valid for the system  $D_2$: $\displaystyle  \rho_{2}({\bf x})=\sum_{j\in I_2} \rho_{2j}({\bf x})$.

Note that 
\begin{eqnarray*}   &  \displaystyle 
\int_{R^{3}} \rho_{1}({\bf x}) d{\bf x}=
\sum_{j\in I_1} \int_{R^{3n_1-3}} \int_{R^{3n_1-3}}| \stackrel{-}{\Phi} (\stackrel{-}{{\bf x}})|^2 d(\stackrel{-}{{\bf x}} \setminus {\bf x}_k) d{\bf x}_k=
\nonumber \\[8pt] &  \displaystyle
=\sum_{j\in I_1} \int_{R^{3n_1}} |\stackrel{-}{\Phi} (\stackrel{-}{{\bf x}})|^2 d\stackrel{-}{{\bf x}} =n_1
\end{eqnarray*}
and $$\displaystyle  \int_{R^{3}} \rho_{2}({\bf x}) d{\bf x}=n_2.$$

\medskip 
We compute the integral
\begin{eqnarray}   &  \displaystyle
 \sum_{j\in I_1 k \in I_2} \int_{R^{3n_1}} \int_{R^{3n_2}} \frac{\stackrel{-}{\Phi}^2 (\stackrel{-}{{\bf x}}) \stackrel{=}{\Phi}^2 (\stackrel{=}{{\bf x}})}{|{\bf x}_k - {\bf x}_j|} d\stackrel{-}{{\bf x}} d\stackrel{=}{{\bf x}} =
 \label{812} \\[8pt] &  \displaystyle
 =\sum_{j\in I_1 k \in I_2} \int_{R^3 \times R^3} \frac{ \displaystyle \int_{R^{3n_1-3}}\stackrel{-}{\Phi}^2 (\stackrel{-}{{\bf x}}) d(\stackrel{-}{{\bf x}} \setminus {\bf x}_k) 
  \int_{R^{3n_2-3}}\stackrel{=}{\Phi}^2 (\stackrel{=}{{\bf x}}) d(\stackrel{=}{{\bf x}_k} \setminus {\bf x}_j)}
{\displaystyle |{\bf x}_k - {\bf x}_j|} d{\bf x}_k d{\bf x}_j =
\nonumber \\[8pt] &  \displaystyle
 =\sum_{j\in I_1 k \in I_2}  \int_{R^3 \times R^3} \frac{\rho_{1j}({\bf x}_k) \rho_{2k}({\bf x}_j)}
{\displaystyle  |{\bf x}_k - {\bf x}_j|} d{\bf x}_k d{\bf x}_j=
\nonumber \\[8pt] &  \displaystyle
 =\sum_{j\in I_1 k \in I_2}  \int_{R^3 \times R^3} \frac{\rho_{1j}({\bf x}) \rho_{2k}({\bf y})}
{\displaystyle  |{\bf x} - {\bf y}|} d{\bf x} d{\bf y}=
\nonumber \\[8pt] &  \displaystyle
 = \int_{R^3 \times R^3} \frac{\displaystyle \sum_{j\in I_1} \rho_{1j}({\bf x}) \sum_{k \in I_2} \rho_{2k}({\bf y})}
{\displaystyle  |{\bf x} - {\bf y}|} d{\bf x} d{\bf y}=
\nonumber \\[8pt] &  \displaystyle
=\int_{R^3 \times R^3} \frac{\displaystyle  \rho_{1}({\bf x}) \rho_{2}({\bf y})}
{\displaystyle  |{\bf x} - {\bf y}|} d{\bf x} d{\bf y}.
\nonumber
\end{eqnarray}

\medskip
We also compute the integral
\begin{eqnarray}   &  \displaystyle
 \sum_{k\in I_1 n \in N_2} \int_{R^{3n_1}} \int_{R^{3n_2}} \frac{\stackrel{-}{\Phi}^2 (\stackrel{-}{{\bf x}}) \stackrel{=}{\Phi}^2 (\stackrel{=}{{\bf x}})Z_n}{|{\bf x}_k - {\bf x}_n|} d\stackrel{-}{{\bf x}} d\stackrel{=}{{\bf x}} 
=  \label{Int}
\\[8pt]   &  \displaystyle
 =\sum_{k\in I_1 n \in N_2}  \int_{R^{3n_2}} \frac{\stackrel{-}{\Phi}^2 (\stackrel{-}{{\bf x}}) Z_n}{|{\bf x}_k - {\bf x}_n|} d\stackrel{-}{{\bf x}} \int_{R^{3n_1}} \stackrel{=}{\Phi}^2 (\stackrel{=}{{\bf x}}) d\stackrel{=}{{\bf x}} =
 \nonumber \\[8pt]   &  \displaystyle
  =\sum_{k\in I_1 n \in N_2}  \int_{R^{3n_2}} \frac{\stackrel{-}{\Phi}^2 (\stackrel{-}{{\bf x}}) Z_n}{|{\bf x}_k - {\bf x}_n|} d\stackrel{-}{{\bf x}} =
  \nonumber \\[8pt]   &  \displaystyle
   =\sum_{k\in I_1 n \in N_2}  \int_{R^{3}} \frac{ \displaystyle \int_{R^{3n_1-3}}\stackrel{-}{\Phi}^2 (\stackrel{-}{{\bf x}}) Z_n d(\stackrel{-}{{\bf x}} \setminus {\bf x}_k)}{|{\bf x}_k - {\bf x}_n|} d{\bf x}_k =
   \nonumber \\[8pt]   &  \displaystyle
 =\sum_{k\in I_1 n \in N_2}  \int_{R^{3}} \frac{ \rho_{1k} ({\bf x}_k) Z_n}{|{\bf x}_k - {\bf x}_n|} d{\bf x}_k =
 \nonumber \\[8pt] &  \displaystyle
 =\sum_{k\in I_1 n \in N_2}  \int_{R^{3}} \frac{ \rho_{1k} ({\bf x}) Z_n}{|{\bf x} - {\bf x}_n|} d{\bf x} =
   \nonumber \\[8pt]   &  \displaystyle
    =\sum_{n \in N_2}  \int_{R^{3}} \frac{\displaystyle \sum_{k\in I_1} \rho_{1k} ({\bf x}) Z_n}{|{\bf x} - {\bf x}_n|} d{\bf x}  =
    \nonumber \\[8pt] &  \displaystyle
    =\sum_{n \in N_2}  \int_{R^{3}} \frac{\rho_{1} ({\bf x}) Z_n}{|{\bf x} - {\bf x}_n|} d{\bf x}.
 \nonumber
\end{eqnarray}

\medskip 
We split the sum into three terms
\begin{eqnarray}   &  \displaystyle
\label{ZZ} 
\sum_{n <  m \in N_1 \cup N_2} \frac{Z_n Z_m}{|{\bf x}_n - {\bf x}_m|}=
 \\[8pt] &  \displaystyle
 =\sum_{n \in N_1  m \in N_1 } \frac{Z_n Z_m}{|{\bf x}_n - {\bf x}_m|}+
 \nonumber \\[8pt] &  \displaystyle
+ \sum_{n \in  N_2  m \cup N_2} \frac{Z_n Z_m}{|{\bf x}_n - {\bf x}_m|}+
 \nonumber \\[8pt] &  \displaystyle
+ \sum_{n \in N_1 \cup N_2  m \in N_1 \cup N_2} \frac{Z_n Z_m}{|{\bf x}_n - {\bf x}_m|}.
 \nonumber
\end{eqnarray}

By using (\ref{812}), (\ref{Int}) and (\ref{ZZ}), we can write (\ref{5}) in the form
\begin{eqnarray}   &  \displaystyle
E=\sum_{k \in I_1 } \int_{R^{3n_1}}  |\nabla_k  \stackrel{-}{\Phi} (\stackrel{-}{{\bf x}})|^2 d\stackrel{-}{{\bf x}}  + \,\,\,\,\,\,\,\,\,\,\,\,\,\,\,\,\,(I)
\label{5aaa} \\[8pt] &  \displaystyle
+ \sum_{k \in I_2 }  \int_{R^{3n_2}} |\nabla_k  \stackrel{=}{\Phi} (\stackrel{=}{{\bf x}})|^2  d\stackrel{=}{{\bf x}} + \,\,\,\,\,\,\,\,\,\,\,\,\,\,\,\,\,\,\,\,(II)
 \nonumber \\[8pt] &  \displaystyle
 + \sum_{j\in I_1 k \in I_1} \int_{R^{3n_1}} \frac{\stackrel{-}{\Phi}^2 (\stackrel{-}{{\bf x}})}{|{\bf x}_k - {\bf x}_j|}d\stackrel{-}{{\bf x}}   + 
\,\,\,\,\,\,\,\,\,\, \,\,\,\,\,\,\,\,\,\,(I)
 \nonumber \\[8pt] &  \displaystyle
 + \sum_{j\in I_2 k \in I_2} \int_{R^{3n_1}} \frac{\stackrel{=}{\Phi}^2 (\stackrel{=}{{\bf x}})}{|{\bf x}_k - {\bf x}_j|}d\stackrel{=}{{\bf x}}   + 
\,\,\,\,\,\,\,\,\,\, \,\,\,\,\,\,\,\,\,\,(II)
 \nonumber \\[8pt] &  \displaystyle
+ \int_{R^3 \times R^3} \frac{\displaystyle  \rho_{1}({\bf x}) \rho_{2}({\bf y})}
{\displaystyle  |{\bf x} - {\bf y}|} d{\bf x} d{\bf y} +
\,\,\,\,\,\,\,\,\,\,\,\,\,\,\,\,\,\,\,\,\,\,\,\,\,(e)
 \nonumber \\[8pt] &  \displaystyle
 + \sum_{j\in I_1 n \in N_1} \int_{R^{3n_1}} \frac{\stackrel{-}{\Phi}^2 (\stackrel{-}{{\bf x}})Z_n}{|{\bf x}_k - {\bf x}_n|} d\stackrel{-}{{\bf x}}  + 
 \,\,\,\,\,\,\,\,\,\,\,\,\,\,\,\,\,\,(I)
 \nonumber \\[8pt] &  \displaystyle
+ \sum_{n \in N_2}  \int_{R^{3}} \frac{\rho_{1} ({\bf x}) Z_n}{|{\bf x} - {\bf x}_n|} d{\bf x} +
\,\,\,\,\,\,\,\,\,\,\,\,\,\,\,\,\,\,\,\,\,\,\,\,\,\,\,\,\,\,\,(e)
 \nonumber  \\[8pt] &  \displaystyle
 + \sum_{j\in I_2 n \in N_2} \int_{R^{3n_1}} \frac{\stackrel{=}{\Phi}^2 (\stackrel{=}{{\bf x}})Z_n}{|{\bf x}_k - {\bf x}_n|} d\stackrel{=}{{\bf x}}  + 
\,\,\,\,\,\,\,\,\,\, \,\,\,(II)
 \nonumber \\[8pt] &  \displaystyle
+ \sum_{n \in I_1}  \int_{R^{3}} \frac{\rho_{2} ({\bf x}) Z_n}{|{\bf x} - {\bf x}_n|} d{\bf x}+ 
\,\,\,\,\,\,\,\,\,\,\,\,\,\,\,\,\,\,\,\,\,\,\,\,\,\,\,\,\,\,\,\,\,\,\,\,(e)
  \nonumber \\[8pt] &  \displaystyle
+ \sum_{n \in N_1  m \in N_1 } \frac{Z_n Z_m}{|{\bf x}_n - {\bf x}_m|}+
 \,\,\,\,\,\,\,\,\,\,\,\,\,\,\,\,\,\,\,\,\,\,\,\,\,\,\,\,\,\,\,\,\,\,\,\,(I)
 \nonumber \\[8pt] &  \displaystyle
+ \sum_{n \in  N_2  m \cup N_2} \frac{Z_n Z_m}{|{\bf x}_n - {\bf x}_m|}+
\,\,\,\,\,\,\,\,\,\,\,\,\,\,\,\,\,\,\,\,\,\,\,\,\,\,\,\,\,(II)
 \nonumber \\[8pt] &  \displaystyle
+ \sum_{n \in N_1 m \in \cup N_2} \frac{Z_n Z_m}{|{\bf x}_n - {\bf x}_m|}.
 \nonumber
\,\,\,\,\,\,\,\,\,\,\,\,\,\,\,\,\,\,\,\,\,\,\,\,\,\,\,\,\,\,\,\,\,(e)
\end{eqnarray}

In (\ref{5aaa}), by collecting the terms corresponding to the system $D_1$ (the terms marked by $I$), the terms  corresponding to the system $D_2$ (the terms marked by $II$) and the remaining terms (those marked by $e$) that correspond to the interaction of the elements of different systems, we can write (\ref{5aaa}) in the form
\begin{eqnarray}   &  \displaystyle
E=E_1(\stackrel{-}{\Phi})+ E_2(\stackrel{=}{\Phi})+ E_e(\rho_{1},\rho_{2}),
\label{E1E2Ee} 
\end{eqnarray}
where 
\begin{eqnarray}   &  \displaystyle
E_1(\stackrel{-}{\Phi})=\sum_{k \in I_1 } \int_{R^{3n_1}}  |\nabla_k  \stackrel{-}{\Phi} (\stackrel{-}{{\bf x}})|^2 d\stackrel{-}{{\bf x}}  +
\label{E1} \\ &  \displaystyle
 + \sum_{j\in I_1 k \in I_1} \int_{R^{3n_1}} \frac{\stackrel{-}{\Phi}^2 (\stackrel{-}{{\bf x}})}{|{\bf x}_k - {\bf x}_j|}d\stackrel{-}{{\bf x}}   +
\nonumber \\ &  \displaystyle
 + \sum_{j\in I_1 n \in N_1} \int_{R^{3n_1}} \frac{\stackrel{-}{\Phi}^2 (\stackrel{-}{{\bf x}})Z_n}{|{\bf x}_k - {\bf x}_n|} d\stackrel{-}{{\bf x}} +
 \nonumber \\[8pt] &  \displaystyle
+ \sum_{n < m \in \cup N_1} \frac{Z_n Z_m}{|{\bf x}_n - {\bf x}_m|}
 \nonumber
\end{eqnarray}
is the total energy of the first quantum system neglecting the interaction with the second quantum system,
\begin{eqnarray}   &  \displaystyle
E_2(\stackrel{=}{\Phi})=\sum_{k \in I_2 } \int_{R^{3n_2}}  |\nabla_k  \stackrel{=}{\Phi} (\stackrel{=}{{\bf x}})|^2 d\stackrel{=}{{\bf x}}  +
\label{E2} \\ &  \displaystyle
 + \sum_{j\in I_2 k \in I_2} \int_{R^{3n_2}} \frac{\stackrel{=}{\Phi}^2 (\stackrel{=}{{\bf x}})}{|{\bf x}_k - {\bf x}_j|}d\stackrel{=}{{\bf x}}   +
\nonumber \\ &  \displaystyle
 + \sum_{j\in I_2 n \in N_2} \int_{R^{3n_2}} \frac{\stackrel{=}{\Phi}^2 (\stackrel{=}{{\bf x}})Z_n}{|{\bf x}_k - {\bf x}_n|} d\stackrel{=}{{\bf x}} +
 \nonumber \\[8pt] &  \displaystyle
+ \sum_{n <  m \in \cup N_2} \frac{Z_n Z_m}{|{\bf x}_n - {\bf x}_m|}
 \nonumber
\end{eqnarray}
is the total energy of the second quantum system neglecting the interaction with the first quantum system, and   
\begin{eqnarray}   &  \displaystyle
E_e(\rho_{1},\rho_{2}) = \int_{R^3 \times R^3} \frac{\displaystyle  \rho_{1}({\bf x}) \rho_{2}({\bf y})}
{\displaystyle  |{\bf x} - {\bf y}|} d{\bf x} d{\bf y} +
 \label{Exx}  \\[8pt] &  \displaystyle
 + \sum_{n \in N_2}  \int_{R^{3}} \frac{\rho_{1} ({\bf x}) Z_n}{|{\bf x} - {\bf x}_n|} d{\bf x} +
 \nonumber  \\[8pt] &  \displaystyle
 + \sum_{n \in N_1}  \int_{R^{3}} \frac{\rho_{2} ({\bf x}) Z_n}{|{\bf x} - {\bf x}_n|} d{\bf x}+
 \nonumber \\[8pt] &  \displaystyle
 + \sum_{ n \in N_1   m \in  N_2} \frac{Z_n Z_m}{|{\bf x}_n - {\bf x}_m|}
 \nonumber
\end{eqnarray}
describes the energy of the mutual interaction of two quantum systems under consideration. It is seen that the energy $E_e$ is an electrostatic-type energy [2].

We see that in (\ref{Exx}) the electrons' charges are ``smeared'' over the corresponding systems in accordance with the prediction by the Schr{\"o}dinger equation for the corresponding system. The electron densities $\rho_{1}({\bf y})$ and $\rho_{2}({\bf y})$ may be deduced in different ways: either computed or determined by an experiment.

The energies $E_1$ (\ref{E1}) and $E_2$ (\ref{E2}) of the first and second ``rigid'' quantum systems are constants (do not depend on the translation and/or rotation of the systems). They do not change when the systems move as rigid bodies. The energy  $E_e$ (\ref{Exx}) depends on the positions of the systems considered as rigid bodies: the positions of the fixed points ${\bf x}_1$ and ${\bf x}_2$ in the systems and the Euler angles [3] ${\omega}_1$ and ${\omega}_2$: $E_e=E_e({\bf x}_1, {\bf x}_2, {\omega}_1, {\omega}_2)$.

Finally, the original problem (\ref{(1aaa)}), (\ref{(1bbb)}) takes the form
\begin{eqnarray}   &  \displaystyle
 E=E_1+E_2+E_e({\bf x}_1, {\bf x}_2, {\omega}_1, {\omega}_2) \to \min, 
 \label{(12e)}
\end{eqnarray}
or, which is the same,
\begin{eqnarray}   &  \displaystyle
 E_e({\bf x}_1, {\bf x}_2, {\omega}_1, {\omega}_2) \to \min,
 \label{(e)}
\end{eqnarray}
where
\begin{eqnarray}   &  \displaystyle
E_e= \int_{G(R^3) \times G(R^3)} \frac{\displaystyle  \rho_{1}({\bf x}) \rho_{2}({\bf y})}
{\displaystyle  |{\bf x} - {\bf y}|} d{\bf x} d{\bf y} +
 \label{Eee}  \\[8pt] &  \displaystyle
 + \sum_{n \in N_2}  \int_{G(R^{3})} \frac{\rho_{1} ({\bf x}) Z_n}{|{\bf x} - G({\bf x}_n)|} d{\bf x} +
 \nonumber  \\[8pt] &  \displaystyle
 + \sum_{n \in N_1}  \int_{G(R^{3})} \frac{\rho_{2} ({\bf x}) Z_n}{|{\bf x} - G({\bf x}_n)|} d{\bf x}\nonumber \\[8pt] &  \displaystyle
 + \sum_{n \in N_1 \cup N_2  m \in N_1 \cup N_2} \frac{Z_n Z_m}{|G({\bf x}_n) - G({\bf x}_m)|}
 \nonumber
\end{eqnarray}
and the mapping $G$ acts as follows: 
\begin{eqnarray}   &  \displaystyle
G=G({\bf x}_1, {\omega}_1)\,:\, \stackrel{-}{{\bf x}}\in R^3  \to S({\omega}_1)\stackrel{-}{{\bf x}}+{\bf x}_1 \in R^3 ,
\label{(mapping)}
\\  &  \displaystyle
G=G({\bf x}_2, {\omega}_2)\,:\, \stackrel{=}{{\bf x}}\in R^3 \to S({\omega}_2)\stackrel{=}{{\bf x}}+{\bf x}_2   \in R^3.
\nonumber
\end{eqnarray}
It describes the movement of the systems as rigid bodies. Here $S({\omega})$ means the rotation with the Euler angles $\omega$. In other words, we arrive at the problem of classical mechanics considering the interaction of two rigid bodies, which is solved with respect to the twelve variables ${\bf x}_1, {\omega}_1, {\bf x}_2, {\omega}_2$. We can fix the degrees of freedom for one of the bodies and arrive at the problem with respect to six variables only.

Our result correlates with the electrostatic (also known as Hellmann-Feinmann) theorem in quantum mechanics [4]. Specific of the problem under consideration is that we consider solid-like systems, which movements are described not by one scalar parameter, but by mapping (\ref{(mapping)})  (this is the grope of movements of rigid body [5]). The basic assumption, which leads to an analogue of the electrostatic  theorem for solid-like (molecule-like) quantum systems, is  the ``rigidity'' of quantum system. 

Formula (\ref{Exx}) means that the first system produces the electric potential
\begin{eqnarray}   &  \displaystyle
P_1({\bf y})= \int_{R^3} \frac{\displaystyle  \rho_{1}({\bf x})} {\displaystyle  |{\bf x} - {\bf y}|} d{\bf x} +
     \sum_{n \in N_1}   \frac{ Z_n}{|{\bf y} - {\bf x}_n|}, 
 \label{Epp} 
\end{eqnarray}
which is the energy of the interaction of the first system with a unit electric charge placed at point ${\bf y}$ outside of the first system. The second system produces a similar electric potential $P_2$.

Probably, the potential (\ref{Epp}) of the quantum system may be measured experimentally.

Let us introduce the functions 
$$\displaystyle Z_1({\bf x})= \sum_{n \in I_1}Z_n \delta({\bf x}-{\bf x}_n), \,\,\,\,
Z_2({\bf x})= \sum_{n \in I_2}Z_n \delta({\bf x}-{\bf x}_n),$$  
where $ \delta({\bf x}) $ means the ``delta''-function with support at the point ${\bf x}=0$ [6]. These functions describe the charge of the nuclei.

We can write the energy $E_e$ (\ref{Exx}) as a bilinear form $L \langle F_1,F_2 \rangle$. Let us introduce the function
$$F_1({\bf x})= \rho_{1}({\bf x})+Z_1({\bf x}),$$ which describes the total distribution of the charge over the first system, and the function $$F_2({\bf x})=\rho_{2}({\bf x})+Z_2({\bf x}),$$ which describes the total distribution of the charge over the second system. Using this notation, we can write $E_e=L \langle F_1,F_2 \rangle$, where
\begin{eqnarray}   &  \displaystyle
L \langle F_1,F_2 \rangle = \int_{R^3 \times R^3} \frac{\displaystyle F_{1}({\bf x}) F_{2}({\bf y})}
{\displaystyle  |{\bf x} - {\bf y}|} d{\bf x} d{\bf y} .
\label{scalar}
\end{eqnarray}

\noindent
{\bf ``Rigid'' quantum systems in external electric field }
\medskip 

\noindent
Practically, it is important to understand the mechanism of interaction of quantum systems in the presence of external fields (in particular, external fields may be used to control quantum systems). 

To take into account the action of an external electric field with potential $V({\bf x})$ on the quantum system under consideration, energy $E(\phi)$ (\ref{(1ccc)}) is supplemented with the term [1]
\begin{eqnarray}   &  \displaystyle
   \label{(1cccsoppl)}  &  \displaystyle \int_{R^{3n_1}\times R^{3n_2}}
 \sum_{j\in I_1 \cup I_2}   V({\bf x}_j)\phi^2({\bf x}) d{\bf x} +  \sum_{n \in N_1 \cup N_2} V({\bf x}_j) Z_n .
 \end{eqnarray} 
 The first sum in (\ref{(1cccsoppl)}) is potential energy of electrons in the field with potential $V(x)$ and the second sum in (\ref{(1cccsoppl)}) is potential energy of nuclei in the field with potential $V(x)$.
 
 Under condition (\ref{(2)}), the first expression (\ref{(1cccsoppl)}) may be transformed to the following way:
\begin{eqnarray}  \label{(001)}  &  \displaystyle
 \sum_{j\in I_1 \cup I_2} \int_{R^{3n_1}} \int_{R^{3n_2}} V({\bf x}_j) \stackrel{-}{\Phi}^2 (\stackrel{-}{{\bf x}}) \stackrel{=}{\Phi}^2 (\stackrel{=}{{\bf x}}) d\stackrel{-}{{\bf x}} d\stackrel{=}{{\bf x}} =
 \label{812aaa}  \\[8pt] &  \displaystyle
 = \sum_{j\in I_1 } \int_{R^{3n_1}}  V({\bf x}_j) \stackrel{-}{\Phi}^2 (\stackrel{-}{{\bf x}}) d\stackrel{-}{{\bf x}}  \int_{R^{3n_2}} \stackrel{=}{\Phi}^2(\stackrel{=}{{\bf x}}) d\stackrel{=}{{\bf x}}+ 
 \nonumber \\[8pt] &  \displaystyle
 +\sum_{j\in I_2 } \int_{R^{3n_1}}  \stackrel{-}{\Phi}^2 (\stackrel{-}{{\bf x}})  d\stackrel{-}{{\bf x}} \int_{R^{3n_2}} V({\bf x}_j) \stackrel{=}{\Phi}^2 (\stackrel{=}{{\bf x}})  d\stackrel{=}{{\bf x}}=
  \nonumber \\[8pt] &  \displaystyle
= \sum_{j\in I_1 } \int_{R^{3n_1}}  V({\bf x}_j) \stackrel{-}{\Phi}^2 (\stackrel{-}{{\bf x}}) d\stackrel{-}{{\bf x}}  + 
\sum_{j\in I_2 }  \int_{R^{3n_2}} V({\bf x}_j) \stackrel{=}{\Phi}^2 (\stackrel{=}{{\bf x}})  d\stackrel{=}{{\bf x}}.
 \nonumber 
 \end{eqnarray}
  In (\ref{(001)}) we use equations (\ref{is1}). We also take into account that $j\in I_1$ corresponds to $\stackrel{-}{{\bf x}}$ and $j\in I_2$ corresponds to $\stackrel{=}{{\bf x}}$.
  
  The following equalities take place:
  \begin{eqnarray}   &  \displaystyle
\sum_{j\in I_1 }  \int_{R^{3}} \int_{R^{3n_1-3}}  V({\bf x}_k) \stackrel{-}{\Phi}^2 (\stackrel{-}{{\bf x}}) d(\stackrel{-}{{\bf x}} \setminus {\bf x}_j) d{\bf x}_j =
\nonumber \\[8pt] &  \displaystyle
=
\sum_{j\in I_1 }  \int_{R^{3}} V({\bf x}_j)  \rho_{1j} ({\bf x}_j) d{\bf x}_j 
=\sum_{j\in I_1 }  \int_{R^{3}} V({\bf x})  \rho_{1j} ({\bf x}) d{\bf x} =
\nonumber \\[8pt] &  \displaystyle
=\int_{R^{3}} V({\bf x}) \sum_{j\in I_1 }  \rho_{1j} ({\bf x}) d{\bf x}=
\int_{R^{3}} V({\bf x})  \rho_{1} ({\bf x}) d{\bf x},
   \nonumber 
   \end{eqnarray}
   Here we use definition (\ref{rho}).
   
   In a similar way, we obtain the equality
\begin{eqnarray}   &  \displaystyle
    \int_{R^{3}} \int_{R^{3n_1-3}}  V({\bf x}_j) \stackrel{=}{\Phi}^2 (\stackrel{=}{{\bf x}}) d(\stackrel{=}{{\bf x}} \setminus {\bf x}_j) d{\bf x}_j =
 \int_{R^{3}} V({\bf x})  \rho_{2} ({\bf x}) d{\bf x}.
    \nonumber
\end{eqnarray}

The second term in (\ref{(1cccsoppl)}) may be written as
\begin{eqnarray}   &  \displaystyle
   \label{(004)}  &  \displaystyle \int_{R^{3n_1}\times R^{3n_2}}
\sum_{n \in N_1 \cup N_2} V({\bf x}_j) Z_n = \sum_{n \in N_1 } V({\bf x}_j) Z_n +\sum_{n \in  N_2} V({\bf x}_j) Z_n .
 \end{eqnarray} 
 
 By using the functions $F_1({\bf x})$ and $F_2({\bf x})$, which describe the total distribution of the charge over the systems, we can write (\ref{(1cccsoppl)}) as
 \begin{eqnarray} \label{VF12}  &  \displaystyle
    \int_{R^{3}} V({\bf x})  F_1({\bf x}) d{\bf x}+ \int_{R^{3}} V({\bf x}) F_2({\bf x}) d{\bf x}.
\end{eqnarray}

Formula (\ref{VF12}) coincides with the formula of classical electrostatics for the sum of potential energies of the system of charges, where $F_1({\bf x})$ and $F_2({\bf x})$ are electric fields with the potential $V({\bf x})$  [2]. 

It means that the ``rigid'' quantum systems react to external electric field like classical electrostatic devices. The principal difference is that this reaction as well as the integrity of ``rigid'' quantum systems is provided with ``quantum forces''.

\medskip \medskip  \medskip  \medskip 
\noindent
{\bf The complementary problems for ``rigid'' quantum systems }
\medskip 

\noindent
The electrostatic model of the interaction of charged bodies was used earlier for the analysis of complementary problems for charged bodies [7, 8].  The existence of charged bodies forming a set of complementary bodies was demonstrated in a mathematical way. The works [7, 8] consider hypothetical stable charged systems interacting one with another in the electrostatic manner, i.e., interacting in the manner described by equation (\ref{scalar}). 

The present paper demonstrates that the interaction of ``rigid'' quantum systems  are candidates to the role of ``real word'' objects which may form a set of complementary bodies working on the basis of the principle proposed in [7, 8].  

An external electric field can influence the complementarity property of bodies interacting in the electrostatic manner. If the bodies forming the system are electro-neutral, a constant electric field cannot influence the complementarity property of the bodies. Actually, in this case expression (\ref{VF12}) is  equal to zero identically. This sounds like a macroscopic electric field cannot influence the complementarity property of microscopic electro-neutral bodies.

\medskip \medskip  \medskip  \medskip 
\noindent
{\bf Conclusions and prospectives}
\medskip 

\noindent
We demonstrated that the interaction of ``rigid'' quantum systems one with the others (also with an external electric field) is similar to the electrostatic interaction of ``rigid''  systems of charges. The point is that a stable ``rigid''  system of charges does not exist, while a stable ``rigid'' quantum systems does. 

This note introduces a new approach to quantum systems (the ``rigid'' ones) based on the idea of electro-mechanical devices known on the macroscopic level [9]. In our approach, the integrity of the ``rigid'' quantum systems is maintained by the ``quantum forces''. The ``quantum forces'' also form the distribution of total charge over systems. But the interaction of ``rigid'' quantum systems is realised in the form that is mathematically equivalent to the electrostatic interaction of rigid bodies possessing a known distribution of charges. This note also indicates a way to control ``rigid'' quantum systems and to develop ``smart'' systems by using known approaches, see, e.g, [10]. 

The present approach is approximate and may be used while the assumption about the ``rigidity'' of our quantum systems is in force. 

We assume that the distribution of the total charge (the functions $F_1({\bf x})$ and $F_2({\bf x})$) over the systems is described by a specific function which reflects the structure of the systems on the level ``electrons -- nuclei'',  i.e., the molecular level.  This means that the notion of ``molecular size'' electro mechanical devices provides an adequate characteristic for the systems under consideration.

\medskip \medskip  \medskip
\noindent
{\bf Acknowledgements}. AGK was supported through the Marie Curie actions FP7 project PIIF2-GA-2008-219690.

\newpage
\noindent
\noindent{\bf References} 
\medskip

1. Cances E., M, Defranceshi M., et al. Computational quantum chemistry: a premier. In:	Computational Chemistry, Ed. C. Le Bris. Norht-Holland, Amsterdam; 2003.

2. Jackson J.D. Classical Electrodynamics. 3rd Ed., Wiley, New York; 1998.

3. Gallavotti G. The Elements of Mechanics. Springer-Verlag, Berlin-Heidelberg-New York; 1983.  

4. Martin R.M. Electronic Structure: Basic Theory and Practical Methods. Cambridge University Press, Cambridge; 2004.

5. Margenau H., Murphy G.M. The Mathematics of Physics and Chemistry, 2 vols. Princeton, NJ: Van Nostrand, 1956-64. 

6. Yosida K. Functional Analysis. Springer, Berlin; 1996.

7. Kolpakov A.A.,  Kolpakov  A.G. Complementarity problems for two pairs of charged bodies. arXiv:1205.5157

8. Kolpakov A.A.,  Kolpakov  A.G. Complementarity problems for electro-neutral charged bodies. arXiv:1207.5142 

9. Kovacs G., D\"uring L., Michel S., Terrasi G. Stacked dielectric elastomer actuator for tensile force transmission. Sensors and Actuators A: Physical. 155(2), 2009.

10. Kolpakov A.A.,  Kolpakov  A.G. Design of ``intelligent'' structures as a discrete optimal problem.	Electr. Notes  Discrete Math., 27, 2006.

\end{document}